\documentclass[12pt]{article}

\usepackage{amsmath}
\usepackage{amsfonts,amssymb,amsthm,overpic}
\makeatletter
\newcommand{\diam}{\mathop{\operator@font diam}}
\usepackage{setspace}
\usepackage{multicol}
\usepackage{multirow}
\usepackage[compact]{titlesec}

\usepackage{epsfig}

\newtheorem{theorem}{Theorem}[section]

\def \be{\begin{equation}}
\def \ee{\end{equation}}
\def \bq{\begin{eqnarray}}
\def \eq{\end{eqnarray}}
\def \beq{\begin{eqnarray*}}
\def \eeq{\end{eqnarray*}}

\newcommand{\cT}{\mathcal{T}}

\begin{document}

\title{\Huge{\textsc{The causal order on the ambient boundary}}}
\author{{\Large\textsc{Ignatios Antoniadis$^{1,2}$\thanks{\texttt{antoniad@lpthe.jussieu.fr}}, Spiros Cotsakis\footnote{On leave from the University of the Aegean, 83200 Samos, Greece.}\,\,$^{3}$\thanks{\texttt{skot@aegean.gr}}}}\\{\Large\textsc{Kyriakos Papadopoulos$^{3}$\thanks{\texttt{Kyriakos.P@aum.edu.kw}},}} \\
$^1$LPTHE, UMR CNRS 7589, Sorbonne Universit\'es, UPMC Paris 6,\\
75005 Paris, France\\
$^2$ Albert Einstein Center for Fundamental Physics, ITP,\\
Bern University,\\ Sidlerstrasse 5 CH-3012 Bern, Switzerland\\
$^{3}$Department of Mathematics, \\ American University of the Middle East\\P. O. Box 220 Dasman, 15453, Kuwait }

\maketitle
\begin{abstract}
\noindent We analyse the causal structure of the ambient boundary, the conformal infinity of the ambient (Poincar\'e) metric. Using topological tools we show that the only causal relation compatible with the global topology of the boundary spacetime is the horismos order. This has important consequences for the notion of time in the conformal geometry of the ambient boundary.
\end{abstract}
\section{Introduction}
Ambient cosmology is an attempt to describe the structure of the universe in the neighborhood of the spacetime singularities predicted by the singularity theorems of general relativity. This idea was introduced in Refs. \cite{ac1,ac2,ac3} where it was shown that singularities are absent, cosmic censorship becomes a valid proposition by construction, and in fact the method of proof of the theorems predicting singularities becomes inapplicable in this framework.

The main motivation of the ambient approach to the problem of singularities in cosmology is that  the infinities that are inherent  in the spacetime metric according to the singularity theorems indicate the necessity of a conformal geometry of metrics, $g\rightarrow \Omega^2 g$, to absorb them,  not a breakdown of general relativity. Any regular metric will be mapped to a different, inequivalent one, so only singularities may give rise to a conformal symmetry.

In the ambient approach, we imagine that our universe $M$ becomes a conformal 4-boundary at infinity of another metric, a conformal metric on an  ambient 5-space $V$. This arises because starting from the cone metric  $g_5=a^2(y)g_4+dy^2$ on $V$, an asymptotic analysis of solutions to the Einstein equations with a fluid source in the ambient space implies a very rich structure at the infinity of $V$: It is possible to express the wrap factor $a(y)$ as a formal power series converging to a smooth function $\sigma(w)$, giving rise to an `ambient' (Poincar\'e) 5-metric $g_+=\omega^{-n}\left(\sigma^2(w)g_4(x^\mu)+dw^2\right),$ $n\in\mathbb{Q}^+$. This allows us to extract a regular metric  $\mathring{g}$ whose restriction to the conformal boundary at infinity $\mathring{g}|_M$ becomes non-degenerate, and due to a subtle asymptotic condition, it gives a conformal structure to the boundary spacetime $M$, the ambient boundary, cf. \cite{ac1}.

Because of the existence of a conformal geometry on the ambient boundary there is no unique metric there, and so all physically important quantities have to be conformally invariant. This places certain constraints on the various causality relations which could in principle be defined there. Another important feature  tightly connected to the whole geometry of the model is the existence of a unique topology on the ambient boundary, the Zeeman topology \cite{ac2}.

In Section 2,  we show that among all possible causal order relations defined on the ambient boundary, only a horismotic one  may be  fully compatible with the Zeeman topology on the boundary. This fact has important consequences for the causality structure of the ambient boundary, a fact we discuss in Section 3.

\section{The interval topology from horismos}
In spacetime geometry, it is standard to introduce three order relations, namely, the
chronological order $\ll$, the causal order $\prec$ and the order horismos $\rightarrow$, and these can be meaningfully extended to  any \emph{event-set}, a set
$(X,\ll,\prec,\rightarrow)$ equipped with all three of these orders having no metric \cite{Penrose-Kronheimer,Penrose-difftopology}. In this context we say that the event $x$ chronologically precedes event $y$, written
$x\ll y$ if $y$ lies inside the future null cone of $x$, $x$ {\em causally precedes} $y$,
$x \prec y$, if $y$ lies inside or on the future null cone of $x$, and  $x$ is at {\em horismos}
with $y$, written $x \rightarrow y$, if $y$ lies on the future null cone of $x$. The chronological and horismotic orders  are anti-reflexive, while the causal order is reflexive. Then the notations $I^+(x) = \{y \in M : x \ll y\},  J^+(x) = \{y \in M : x \prec y\}$ will be used for the chronological and the causal futures  of $x$ respectively (and with a minus instead of a plus sign  for the pasts), while the future null cone of $x$ will be denoted by $\mathcal{N}^+(x)\equiv\partial J^{+}(x)= \{y \in M : x \rightarrow y\}$, and dually for the null past of $x$, cf. \cite{Penrose-difftopology}. Our metric signature is timelike, $(+,-,-,-)$.

The above definitions of  futures and the pasts of a set can be trivially extended to the situation of any partially ordered set $(X,<)$, and the so-called \emph{orderability  problem} is concerned with the conditions under which the topology $\cT_<$ induced
by the order $<$  is equal to some given topology $T$ on $X$, and conversely,  the possible topologies induced on it by any given order, cf. \cite{Compendium}, \cite{Good-Papadopoulos}, \cite{Orderability-Theorem}.

In a purely topological context this is usually done by passing to the so-called upper (i.e., future) and lower (i.e., past)
sets which in turn lead to the  future  and past  topologies (see \cite{Compendium}).
A subset $A \subset X$ is a {\em past set} if $A = I^-(A)$ and dually for the future. Then, the {\em future topology} $\cT^+$ is generated
by the subbase $\mathcal{S}^+ = \{X \setminus I^-(x) : x \in X\}$
and the {\em past topology} $\cT^-$ by $\mathcal{S}^- = \{X \setminus I^+(x)  : x \in X\}$.
The {\em interval topology} $\cT_{in}$ on $X$ then consists of basic sets which are finite intersections
of subbasic sets of the past and the future topologies. This is in fact the topology that fully characterizes a given order of the poset $X$, since it is  solely generated by the order properties and nothing else.

For reasons explained in the Introduction, we will particularly focus on the development of the orderability problem in the context of the {\em Zeeman fine topology} $F$,  defined as the finest topology with the property that on $M$ it induces the $3$-dimensional Euclidean topology on every space axis and the $1$-dimensional Euclidean topology on every time axis, cf. \cite{Zeeman1,gobel}. We shall use  $M^E$ to denote  $M$ equipped with the $4$-dimensional, standard (manifold),  locally-Euclidean topology, and $M^F$ to mean
that $M$ is equipped with the Zeeman fine topology.

In fact, we shall consider a somewhat subtler version of the fine topology, namely, the topology given by the basic
sets of the form $B_{\epsilon}^F(x)$, where $x \in M$ and $\epsilon > 0$. This is  coarser than $M^F$ but finer than $M^E$ (see \cite{Zeeman1}). We denote this topology by $Z$, and call it \emph{the} Zeeman topology. In
particular, $B_{\epsilon}^F(x) = \bigl(B_{\epsilon}^E(x) - N(x)\bigr) \cup \{x\}$, where $B_{\epsilon}^E(x)$ is the Euclidean open ball and $\mathcal{N}(x)$ the nullcone at $x$. In other words, we remove $\mathcal{N}(x)$ and put back only the point $x$. Obviously, the sets $B_{\epsilon}^F(x)$ are not Euclidean-open.

What is the relation between the Zeeman topology $Z$  and that induced from  horismos?
From the definition of horismos,  we have that $x \rightarrow y$, if $x \prec y$ but not $x \ll y$. For the construction of the interval topology $T_{\rightarrow}$  induced from $\rightarrow$, we first consider the complements of
the sets $\mathcal{N}^{\pm}(x)$, that is the future and past nullcones of $x$,
\be
M \setminus \mathcal{N}^{+}(x) = \{y \in M: y \ll x \textrm{ or } x \prec y \textrm{ or } y \nprec x\},
\ee
 and dually for $M \setminus \mathcal{N}^{-}(x)$.
Then we observe that the intersection of the subbasic sets $M \setminus \mathcal{N}^{+}(x)$ and $M \setminus \mathcal{N}^{-}(x)$
gives a Zeeman basic set  because for each event $x$, $M \setminus \mathcal{N}^{+}(x) \cap M \setminus \mathcal{N}^{-}(x)$ gives a Euclidean neighbourhood
with the nullcone removed, but $x$ is kept since $\rightarrow$ is anti-reflexive. We have thus shown the following result.
\begin{theorem}\label{main}
The order horismos induces the Zeeman topology on $M$.
\end{theorem}
It is interesting that neither the chronological order $\ll$ nor the causal order $\prec$ induce a topology equal to the Zeeman topology.
In particular, for an event $x$, the interval topology $T_{\ll}$ from $\ll$ gives basic  sets containing $x$, the light cone of $x$ as well as all events $y$ where $y \nprec x$, but does not include any event inside the future or the past nullcone of $x$. In a similar fashion, the interval topology $T_\prec$ from $\prec$ gives basic open sets containing $y$ and all events $x$ such that $x \nprec y$, without the light and time cones of $y$.

Further, by its construction, the Alexandrov topology \cite{Penrose-difftopology} does not agree with the Zeeman topology: For every event in spacetime, an Alexandrov-open set is the union of sets of the form $<u,v> = \{w: u \ll w \ll v\}$),  different from $T_\ll$.

Also any order different than horismos fails to give the Zeeman topology. This is even true for the horismotic versions of the causality and chronology relations constructed  by using only the properties of the horismos order as in \cite{Penrose-Kronheimer}, as we show next.

Starting from the horismos relation $\rightarrow$,  we may construct orders similar to chronology and causality as follows, cf. \cite{Penrose-Kronheimer}:
 $x$ is \emph{chronologically related to $y$ through horismos}, $x \prec^{\mathfrak{U}}y$, if and only if there exists a finite sequence $u_1,\cdots,u_n$, such that
$x=u_1 \rightarrow \cdots \rightarrow u_n=y$, while  $x$ is \emph{causally related to $y$ through horismos}, $x \ll^{\mathfrak{U}}y$ if and only if $x \prec^{\mathfrak{U}}y$
but not $x \rightarrow y$.
Then we get
subbasic sets which are complements of sets $\uparrow\{x\}=\{y: x \prec^{\mathfrak{U}}y\}$, and similarly for the set
$\downarrow\{x\}$. The  sets
$\uparrow\{x\}$, for example, take the form $\{x:\,\exists\,u_1=x,\cdots,u_n=y, \textrm{with}\,\, x\rightarrow u_2 \rightarrow \cdots \rightarrow u_{n-1} \rightarrow y\}$,
and so their complements will be unions of Zeeman-open balls. Similarly for $\downarrow\{x\}$, and of course when we  intersect unions of Zeeman-open balls we get Zeeman-open sets. This consequently does not guarantee that we will get the whole topology $Z$.

We have therefore shown the following result.
\begin{theorem}\label{relation_between_orders}
The topologies from the chronological and causal orders do not generate the Zeeman topology $Z$.
In addition, the interval topologies $T_{\prec^{\mathfrak{U}}}$ and $T_{\ll^{\mathfrak{U}}}$ are subsets of the Zeeman topology $Z$.
\end{theorem}

\section{Discussion}
In this paper we have analyzed the causality structure on the ambient boundary by purely topological considerations. We have shown that among all possible causal relations on $M$, the only one compatible with the topology of the ambient boundary is the horismotic relation. This is the causal order permitted by the conformal geometry on the boundary.

In fact, due to the presence of a conformal (homothetic) symmetry on the ambient boundary, no other structure except the null structure may have any physical significance. In particular, there can be no objective notion of a timelike or spacelike curve on the boundary, and this manifests itself here in the compatibility of only the horismotic  (null) relation to the underlying topology of the ambient boundary amongst all other possible order relations. As it was already noted in \cite{ac1}, this is related to the idea of masslessness inherent in the ambient boundary.

It is interesting that this gives rise to a novel resolution of the problem of time in the present context. Although time continues to be one of the coordinates in the ambient boundary, it loses any relation to the idea of causality as it is known in general relativity, and in fact, in the conformal infinity of the ambient 5-space only null relations are permitted, compatible with the ambient boundary conformal geometry and the topology from horismos. In a sense, time is not felt by any ambient boundary observer, while massless particles like photons and gravitons move on the null cone.

An important issue that then arises is how to define the idea of dynamical evolution on the ambient boundary when only horismotic relations exist in it. The fact that we may in fact do so is related to the ambient space and the extra dimension. Also we may use the results of the present paper as a means to probe the topology of the ambient space. We shall return to these issues in a future publication.


\begin{thebibliography}{99}
\bibitem{ac1} I. Antoniadis, S. Cotsakis, \emph{Eur. Phys. J. C. } \textbf{75:35} (2015) 1-12,  \textsc{CERN-PH-TH}/2014-176, \texttt{arXiv:1409.2220}.
\bibitem{ac2} I. Antoniadis, S. Cotsakis, \emph{Mod. Phys. Lett. A.} \textbf{30} (2015) 1550161, \texttt{arXiv:1505.04737}.
\bibitem{ac3} I. Antoniadis, S. Cotsakis, \emph{The large-scale structure of the ambient boundary}, \texttt{arXiv:1512.09216}.
\bibitem{Penrose-Kronheimer} E. H. Kronheimer and R. Penrose, \emph{Proc. Camb. Phil. Soc.} \textbf{63} (1967) 481-501.

\bibitem{Penrose-difftopology} R. Penrose, \emph{Techniques of Differential Topology in Relativity}
(CBMS-NSF Regional Conference Series in Applied Mathematics, 1972).

\bibitem{Compendium} G. Gierz \emph{et al},  \emph{A compendium of continuous lattices} (Springer-Verlag, 1980).
\bibitem{Good-Papadopoulos} C. Good, K. Papadopoulos,  \emph{Topology Appl} \textbf{159} (2012) 1565-1572.
\bibitem{Orderability-Theorem} K. Papadopoulos, {\sl On the Orderability Problem and the Interval Topology}, In: \emph{Topics in Mathematical Analysis and Applications},  T. Rassias and L. Toth Eds., (Optimization and Its Applications Springer Series, Springer-Verlag, 2014).

\bibitem{Zeeman1} E. C. Zeeman, \emph{Topology} \textbf{6} (1967) 161-170.
\bibitem{gobel} R. Gobel, \emph{Comm. Math. Phys.} \textbf{46} (1976) 289-307.
\bibitem{Zeeman2} E. C. Zeeman, \emph{J. Math. Phys.} \textbf{5 }(1964) 490-493.

\end{thebibliography}
\end{document}